\documentclass[journal]{IEEEtran}

\usepackage{algorithm}
\usepackage{algorithmicx}
\usepackage[noend]{algpseudocode} % http://ctan.org/pkg/algorithmicx

\usepackage{titlesec} %

\titleformat{\subsubsection}[block]{\normalsize\bfseries}{\arabic{subsection}-\alph{subsubsection}}{1em}{}[]
\titleformat{\paragraph}[block]{\small\bfseries}{[\arabic{paragraph}]}{1em}{}[]

\usepackage{acro}
\usepackage{graphicx}
\usepackage{float}
\usepackage{xcolor}
\usepackage{amsmath}
\usepackage{spconf}

\renewcommand\IEEEkeywordsname{Keywords}

\DeclareAcronym{FCN}{
short=FCN,
long=fully convolutional network,
}

\begin{document}
\title{ASIST: ANNOTATION-FREE SYNTHETIC INSTANCE SEGMENTATION AND TRACKING BY ADVERSARIAL SIMULATIONS}

% \author{Quan~Liu,
%         Isabella M. Gaeta,
%         Mengyang Zhao,
%         Ruining Deng,
%         Aadarsh Jha,
%         Bryan A. Millis,
%         Anita Mahadevan-Jansen,
%         Matthew J. Tyska,
%         Yuankai~Huo$^{\star}$~\thanks{$^{\star}$ Corresponding author. \\E-mail address: yuankai.huo@vanderbilt.edu
%         \\Postal address: Vanderbilt University, Computer Science, Nashville, TN, USA 37215 }
% }

\name{\begin{tabular}{c}
Quan Liu $^{\star}$ \quad Isabella M. Gaeta $^{\dagger}$ \quad Mengyang Zhao $^{\mathsection}$ \quad Ruining Deng $^{\star}$ \quad Aadarsh Jha $^{\star}$ 
\\ \textit{Bryan A. Millis} $^{\dagger}$ \quad \textit{Anita Mahadevan-Jansen} $^{\mathparagraph}$ \quad \textit{Matthew J. Tyska} $^{\dagger}$ \quad \textit{Yuankai Huo} $^{\star}$ \thanks{Corresponding author: Yuankai Huo.}
% \thanks{Email: quan.liu@vanderbilt.edu (Quan Liu)} \thanks{isabella.m.gaeta@vanderbilt.edu (Isabella M. Gaeta)} \thanks{mengyang.zhao@tufts.edu (Mengyang Zhao)} \thanks{r.deng@vanderbilt.edu (Ruining Deng)} \thanks{aadarsh.jha@vanderbilt.edu (Aadarsh Jha)} \thanks{bryan.a.millis@vanderbilt.edu (Bryan A. Millis)} \thanks{anita.mahadevan-jansen@vanderbilt.edu (Anita Mahadevan-Jansen)} \thanks{matthew.tyska@vanderbilt.edu (Matthew J. Tyska)} 
\thanks{Email address: yuankai.huo@vanderbilt.edu (Yuankai Huo)}
\end{tabular}}

\address{$^{\star}$ Vanderbilt University, Computer Science, Nashville, TN, USA 37215  \\
$^{\dagger}$ Vanderbilt University, Cell and Developmental Biology, Nashville, TN, USA 37215 \\
$^{\mathsection}$ Tufts University, Computer Science, Medford, MA, USA 02155 \\
$^{\mathparagraph}$ Vanderbilt University, Biomedical Engineering, Nashville, TN, USA 37215\\
}

\maketitle

\begin{abstract}
\textbf{Background}: The quantitative analysis of microscope videos often requires instance segmentation and tracking of cellular and subcellular objects. The traditional method consists of two stages: (1) performing instance object segmentation of each frame, and (2) associating objects frame-by-frame. Recently, pixel-embedding-based deep learning approaches these two steps simultaneously as a single stage holistic solution. Pixel-embedding-based learning forces similar feature representation of pixels from the same object, while maximizing the difference of feature representations from different objects. However, such deep learning methods require consistent annotations not only spatially (for segmentation), but also temporally (for tracking). In computer vision, annotated training data with consistent segmentation and tracking is resource intensive, the severity of which is multiplied in microscopy imaging due to (1) dense objects (e.g., overlapping or touching), and (2) high dynamics (e.g., irregular motion and mitosis). Adversarial simulations have provided successful solutions to alleviate the lack of such annotations in dynamics scenes in computer vision, such as using simulated environments (e.g., computer games) to train real-world self-driving systems.

\textbf{Methods}: In this paper, we propose an annotation-free synthetic instance segmentation and tracking (ASIST) method with adversarial simulation and single-stage pixel-embedding based learning. 

\textbf{Contribution}: The contribution of this paper is three-fold: (1) the proposed method aggregates adversarial simulations and single-stage pixel-embedding based deep learning; (2) the method is assessed with both the cellular (i.e., HeLa cells) and subcellular (i.e., microvilli) objects; and (3) to the best of our knowledge, this is the first study to explore annotation-free instance segmentation and tracking study for microscope videos.

\textbf{Results}: The ASIST method achieved an important step forward, when compared with fully supervised approaches: ASIST shows 7\% to 11\% higher segmentation, detection and tracking performance on microvilli relative to fully supervised methods, and comparable performance on Hela cell videos.

\end{abstract}

% Note that keywords are not normally used for peer review papers.

% \fi
\IEEEpeerreviewmaketitle

\section{Introduction}
\IEEEPARstart{C}{apturing} cellular and subcellular dynamics through microscopy approaches helps domain experts in characterizing biological processes~\cite{meenderink2019actin} in a quantitative manner, leading to advanced biomedical applications (e.g., drug discovery)~\cite{arbelle2018probabilistic}.

Numerous image processing approaches have been proposed for precise instance object segmentation and tracking. Most of the previous solutions~\cite{al2018deep,korfhage2020detection,van2016deep} follow a similar “two-stage” strategy: I. segmentation on each frame, and II. frame-by-frame association across the video. In recent years, a new family of “single-stage” algorithms was enabled by cutting-edge pixel-embedding based deep learning~\cite{zhao2020faster,payer2018instance}. Such methods enforce the spatiotemporally consistent pixel-wise feature embedding for the same cellular or subcellular objects across video frames. However, pixel-wise annotations require spatial (segmentation) and temporal (tracking) consistency. Such labeling efforts are typically expensive, and potential unscalable, for microscope videos due to I. dense objects (e.g., overlapping or touching), and II. high dynamics (e.g., irregular motion and mitosis). Therefore, better learning strategies are desired beyond the current human annotation based supervised learning.

Adversarial simulation has provided a scalable option to create realistic synthetic environments without extensive human annotations. Particularly striking examples include a) using computer games such as Grand Theft Auto to train self-driving deep learning models~\cite{johnson2016driving}, b) using a simulation environment Gazebo to train robotics~\cite{zamora2016extending}, and c) using a SUMO simulator to train traffic management artificial intelligence (AI)~\cite{kheterpal2018flow}.

\begin{figure}[t]
\begin{center}
\includegraphics[width=1\linewidth]{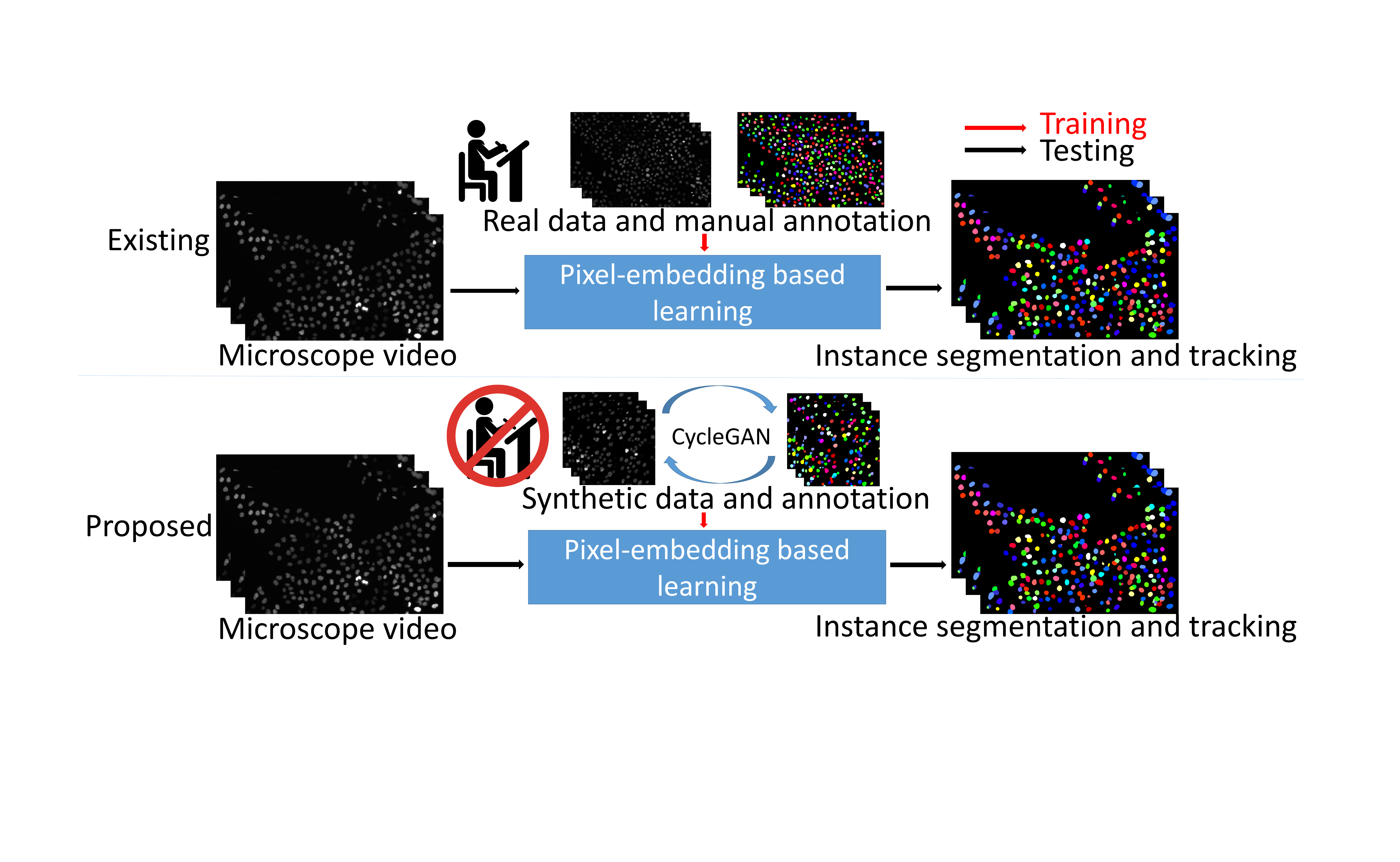}
\end{center}
\caption{The upper panel shows the existing pixel-embedding deep learning based single-stage instance segmentation and tracking method, which is trained by real microscope video and manual annotations. The lower panel presents our pro-posed annotation-free ASIST method, with synthesized data and annotations from adversarial simulations.}
\label{fig:fig1}
\end{figure}

In this paper, we propose an annotation-free synthetic instance segmentation and tracking (ASIST) method with adversarial simulation and single-stage pixel-embedding based learning. Briefly, the ASIST framework consists of three major steps: I. unsupervised image-annotation synthesis, II. video and temporal annotation synthesis, and III. pixel-embedding based instance segmentation and tracking. As opposed to traditional manual annotation-based pixel embedding deep learning, the proposed ASIST method is annotation-free (Fig.\ref{fig:fig1}).

\begin{figure}
\begin{center}
\includegraphics[width=1\linewidth]{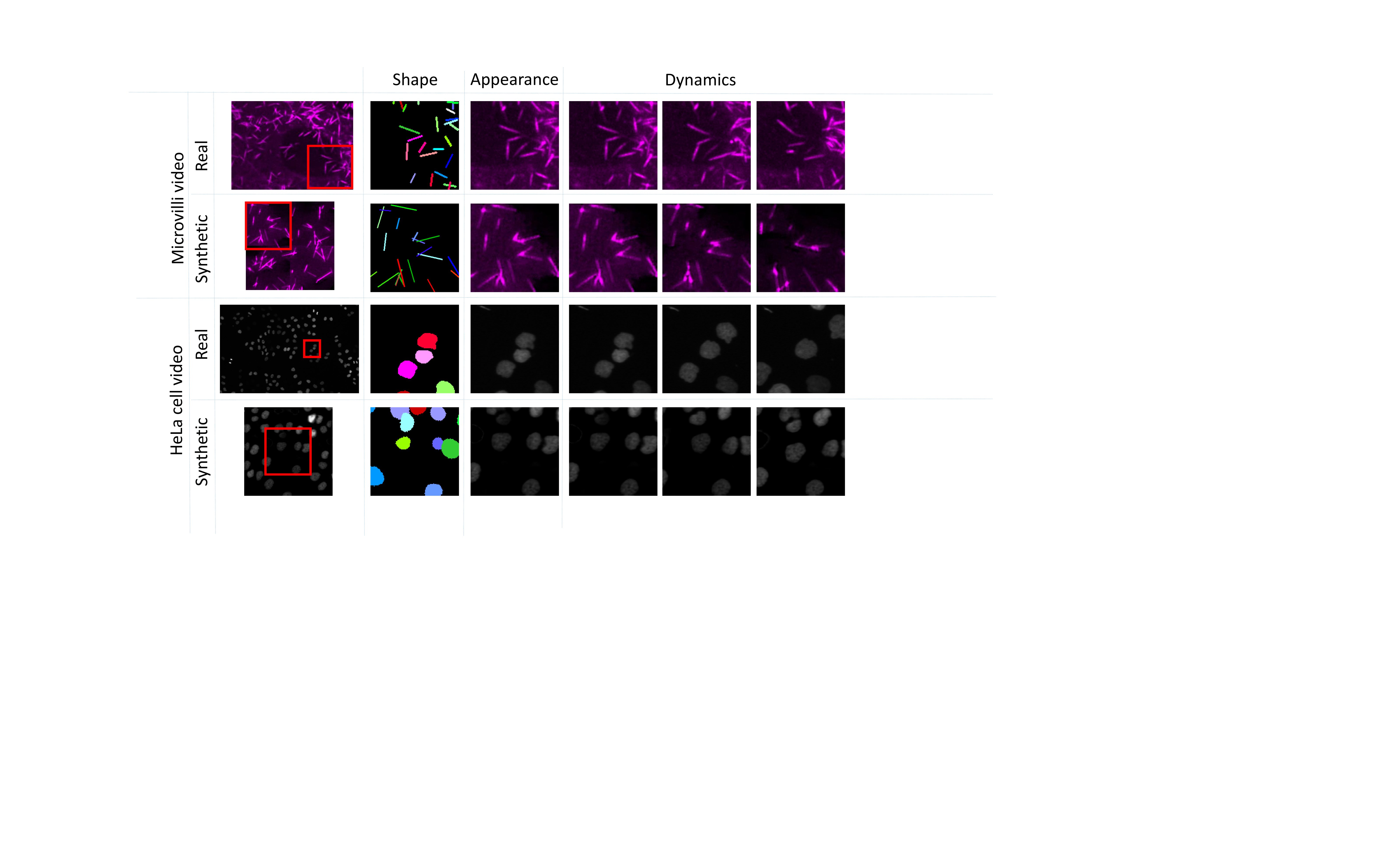}
\end{center}
\caption{Real and synthetic video of Hela cell and microvilli consisting of three aspects: shape, appearance and dynamics. The "shape" is defined as the underlying shape of the manual annotations. The "appearance" is defined by the various appearances of objects. The "dynamics" indicates the mitigation of cellular and subcellular objects.}
\label{fig:fig2}
\end{figure}

To achieve the annotation-free solution, we simulated cellular or subcellular structures with three important aspects: shape, appearance and dynamics (Fig.\ref{fig:fig2}). To evaluate our proposed ASIST method, microscope videos of both cellular (i.e., HeLa cell videos from ISBI Cell Tracking Challenge~\cite{mavska2014benchmark,ulman2017objective}) and subcellular (i.e., microvilli videos from in house data) objects were included in this study. The HeLa cell videos have larger shape variations compared with microvilli videos. From the results, our ASIST method achieved promising accuracy compared with fully supervised approaches. 

In summary, this paper has three major contributions: 
\begin{itemize}
%   \item[$\cdot$ bla1] item 1
  \item We propose the ASIST annotation-free framework, aggregating adversarial simulations and single-stage pixel embedding based deep learning.
  \item We propose a novel annotation refinement approach to simulate shape variations of cellular objects, with circles as a middle representation.
  \item To our best knowledge, our proposed approach is the first annotation-free solution for single-stage pixel embedding deep learning based cell instance segmentation and tracking.
\end{itemize}

% This work extends our conference submission~\cite{liu2020asist} with new efforts: (1) our ASIST method is presented with more details, (2) our ASIST method is validated on a new HeLa cell dataset, and (3) we proposed the annotation refinement to model more complex shape variations compared with~\cite{liu2020asist}.

This research was supported by Vanderbilt Cellular, Biochemical and Molecular Sciences Training Grant 5T32GM008554-25, the NIH NIDDK National Research Service Award F31DK122692, NIH Grant R01-DK111949 and R01-DK095811.

\section{Related Work}

\subsection{Image synthesis}

The simplest approach to synthesize new images is to perform image transformations, which includes flipping, rotation, resizing, and cropping. Such synthetic images improved the accuracy of image classification upon benchmark datasets~\cite{simard2003best} by enlarging dataset with synthetic images. Another study~\cite{drozdzal2018learning} improved the accuracy of image segmentation (Dice similarity coefficient) with synthetic images by applying data augmentation approaches like random sheering and rotation.

A method that is more complex than image transformations are generative adversarial networks (GAN)~\cite{goodfellow2014generative}, which open a new window of synthesizing highly realistic images, and have been widely used in different computer vision and biomedical imaging applications. For instance, GAN has synthesized retinal images to map retinal images to binary retinal vessel trees~\cite{costa2017towards}. The synthetic images can be generated from random noise~\cite{8419363} with geometry constraints~\cite{zhuang2020geometrically}, and even in high dimensional space~\cite{liu2018decompose}. To tackle the limitations of needing paired training data requirements, CycleGAN~\cite{zhu2017unpaired} was proposed to further advance the GAN technique to broader applications. CycleGAN has shown promise in cross-modality synthesis~\cite{huo2018adversarial} and microscope image synthesis~\cite{ihle2019unsupervised}. DeepSynth~\cite{dunn2019deepsynth} demonstrated that CycleGAN can be applied to 3D medical image synthesis.

\subsection{Microscope image segmentation and tracking}

Historically, early approaches utilized intensity-based thresholding to segment a region of interest (ROI) from the background. Ridler et al.~\cite{ridler1978picture} use a dynamic updated threshold to segment an object based on the mean intensity of the foreground and the background. Otsu et al.~\cite{otsu1979threshold} set a threshold by minimizing variance of the intraclass. To avoid the sensitivity to all image pixels, Pratt et al.~\cite{pratt2007digital} proposed growing a segmented area from a point, determined by texture similarity. Based on rough annotations, energy functions can be abstracted to segment images by minimizing the aforementioned energy function~\cite{kass1988snakes}. Among such methods, the watershed segmentation approaches are arguably the most widely used methods for intensity based cell image segmentation~\cite{kornilov2018overview}..

Object tracking on microscope videos is challenging due to the complex dynamics and vague instance boundaries when at cellular or subcellular resolutions. Gerlich et al.~\cite{gerlich2003quantitative} used optical flow from microscope videos to track cell motion. Ray et al.~\cite{ray2004motion} tracked leukocytes by computing gradient vectors of cell motions based on active contours. Sato et al.~\cite{sato1997automatic} designed orientation-selective filters to generate spatio-temporal information by enhancing the motion of cells.~\cite{de1998gastrin,de1999vitro} also tracked cell motion by applying spatiotemporal analysis on microscope videos.

Recent studies have employed machine learning, especially deep learning approaches, for instance cell segmentation and tracking. Jain et al.~\cite{jain2007supervised} showed superior performance of a well-trained convolutional network. Baghli et al.~\cite{baghli2020plasma} achieved 97\% prediction accuracy by employing supervised machine learning approaches. To avoid relying on image annotation, Yu et al.~\cite{yu2018unsupervised} trained a Convolutional Neural Network without annotation to track large scale fibers in images of material acquired via microscope techniques. However, to the best of our knowledge, no existing studies have investigated the challenging problem of quantifying cellular and subcellular dynamics with pixel-wise instance segmentation and tracking with embedding based deep learning.

\section{Methods}

 \begin{figure*}[h]
\begin{center}
\includegraphics[width=0.85\linewidth]{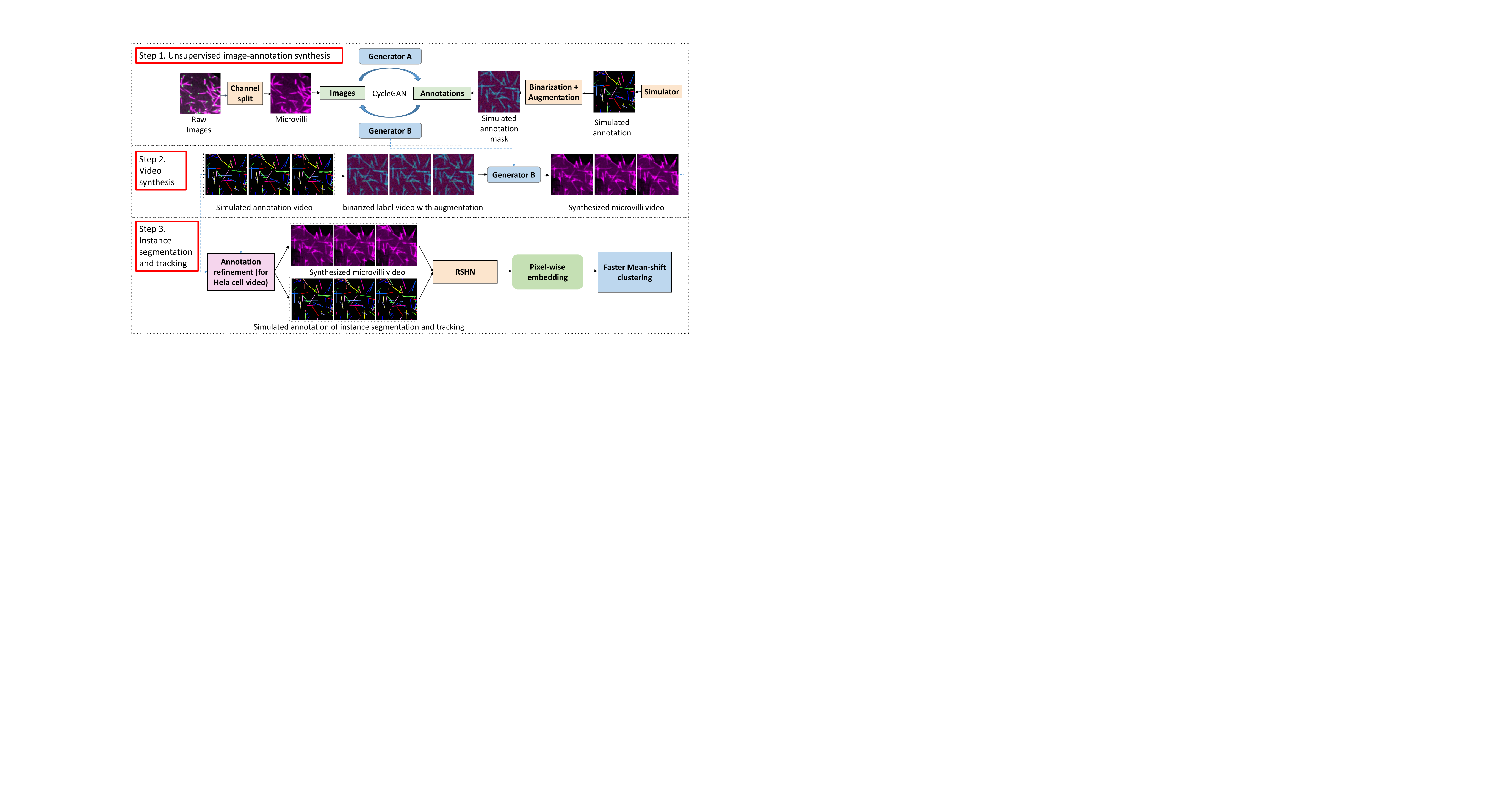}
\end{center}
   \caption{This figure shows the proposed ASIST method. First, CycleGAN based image-annotation synthesis is trained using real microscope images and simulated annotations. Second, synthesized microscope videos are generated from simulated annotation videos. Last, an embedding based instance segmentation and tracking algorithm is trained using synthetic training data. For HeLa cell videos, a new annotation refinement step is introduced to capture the larger shape variations.}
\label{fig:fig3}
\end{figure*}

Our study has three steps: unsupervised image-annotation synthesis, video synthesis and instance segmentation and tracking (Fig.\ref{fig:fig3}). 

\subsection{Unsupervised image-annotation synthesis}

The first step is to train a CycleGAN based approach~\cite{CycleGAN2017} to directly synthesize annotations from microscope images, and vice versa. Compared with the tasks in computer vision, the objects in microscope images are often repetitive with more homogeneous shapes. Therefore, with knowledge of shapes associated with microvilli (stick-shaped) and HeLa cell nuclei (ball-shaped), we randomly generate fake annotations with repetitive sticks and circles to model the shape of microvilli and HeLa cells, respectively. When we train the CycleGAN on microvilli images, we clean the green marks on raw microvilli images which is EPS8 protein by splitting channel of RGB images. The network structure, training process and parameters follows~\cite{liu2020gan}. The generator in CycleGAN consists of an encoder, transformer and decoder. We used ResNet~\cite{he2016deep} with 9 residual blocks as the encoder in both Generator A and Generator B in the deep learning architecture. We have tried to employ U-Net
~\cite{ronneberger2015u} as the encoder as well, suggested by~\cite{liu2020gan}. Based on the our experience, ResNet generally has superior performance compared with U-Net. As a result, the ResNet is employed as the generator through all experiments in this paper. 

\subsection{Video synthesis}
Using an annotation-to-image generator (marked as Generator B) from the above CycleGAN model, synthetic intensity images can be generated from simulated annotations. Since a video dataset represents a compilation of image frames, we extend the utilization of the trained Generator B from “annotation-to-image” to ”annotation frames-to-video”. Briefly, simulated annotation videos are generated by our annotation simulator with variations in shape and dynamics. Then, each annotation video frame is used to generate a synthetic microscope image frame. After repeating such a process for the entire simulated annotation videos, synthetic microscope video is achieved for microvilli and HeLa cells, respectively.

\subsubsection{Microvilli simulation}
\begin{figure*}
\begin{center}
\includegraphics[width=0.8\linewidth]{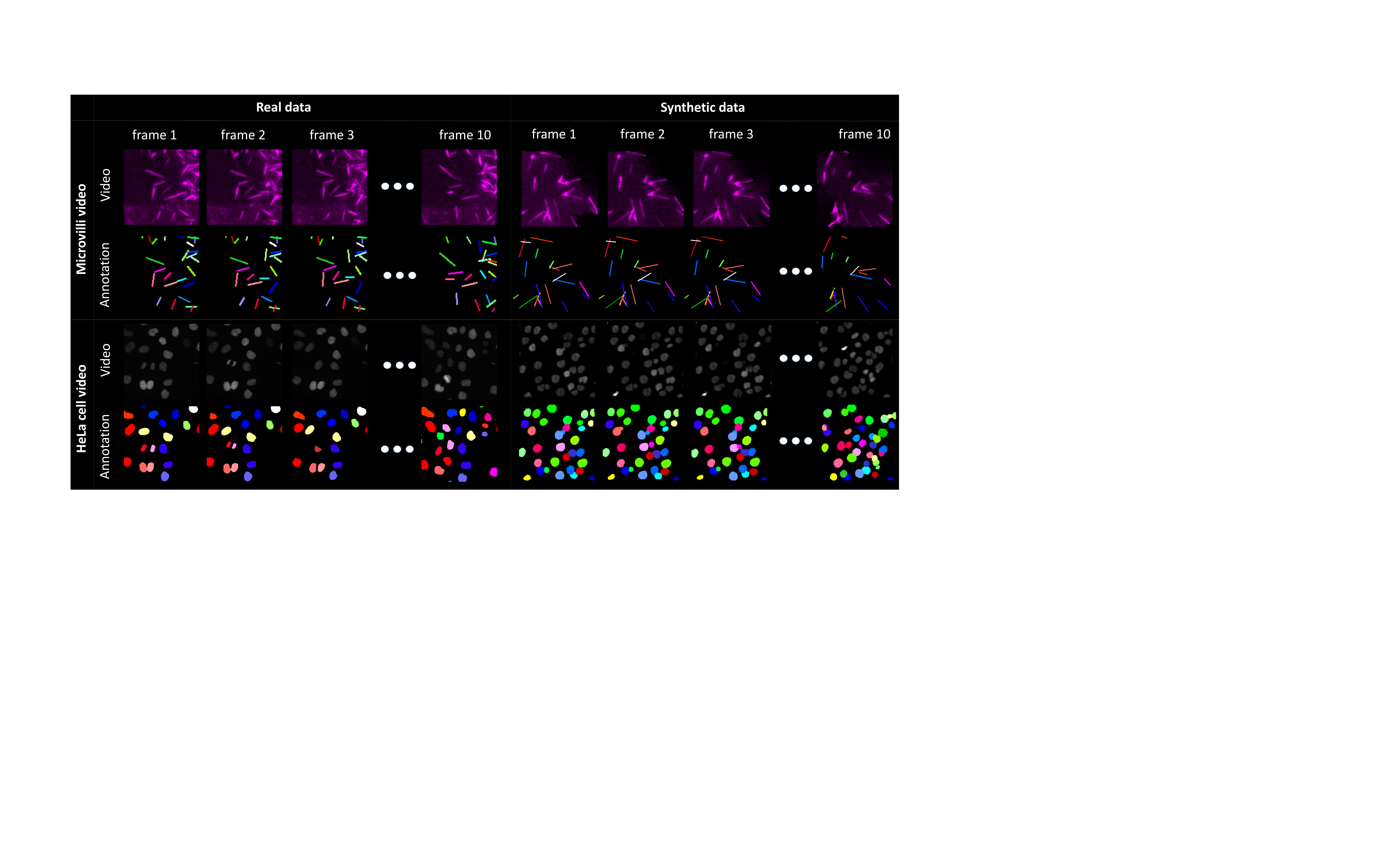}
\end{center}
   \caption{The left panel shows real microscope videos as well as manual annotations. The right panel presents our synthetic videos and simulated annotations.}
\label{fig:fig4}
\end{figure*}

As shown in Fig.\ref{fig:fig4}, we model the shape of microvilli as sticks (narrow
 rectangles) to simulate microvilli videos. The simulated microvilli annotation videos are
 determined by the following operations:
 
\noindent\textbf{Object number}: Different numbers of objects are evaluated when simulating microvilli videos. The details are presented in $\mathsection$\textbf{Experimental design}.

\noindent\textbf{Translation}: Instance annotations are translated by 1 pixel at 50\% probability.

\noindent\textbf{Rotation}: Each instance label is randomly rotated by 1 degree at 50\% probability.

\noindent\textbf{Shortening/Lengthening}: Each object has 50\% probability to become longer or shorter by 1 pixel. Each object can only become longer or shorter across the video.

\noindent\textbf{Moving in/out}: To simulate the instance moving in and out from the video scope, we generate frames in larger size (550 $\times$ 550 pixels) and center-cropped into the target size (512 $\times$ 512 pixels). 

% \begin{figure}
% \begin{center}
% \includegraphics[width=1\linewidth]{Figures/fig3.pdf}
% \end{center}
%   \caption{This figure shows synthesized microvilli videos and simulated label images. The different panels indicate the different number of simulated objects.}
% \label{fig:fig3}
% \end{figure}

\subsubsection{HeLa cell simulation}

\begin{figure}
\begin{center}
\includegraphics[width=0.75\linewidth]{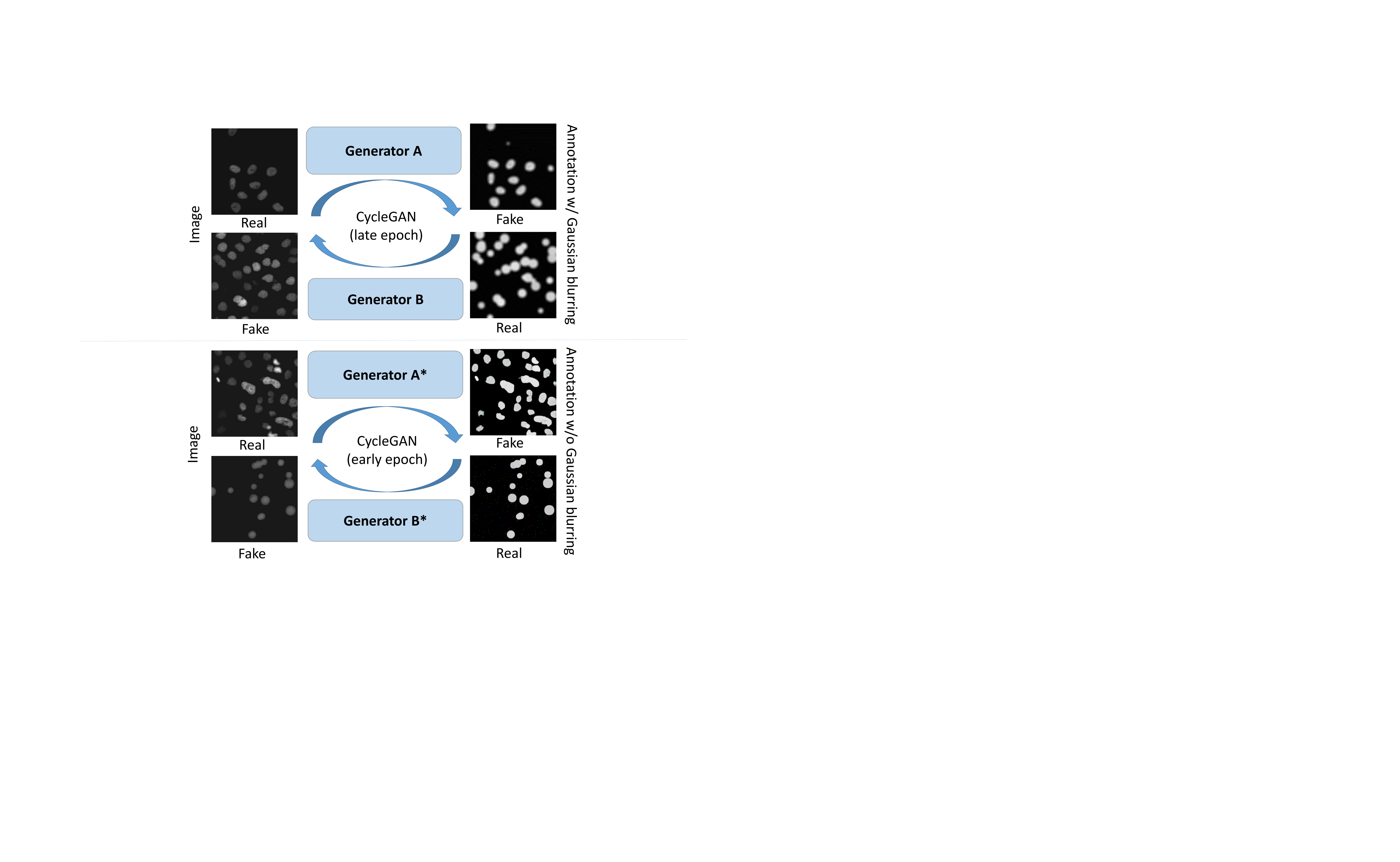}
\end{center}
   \caption{The upper panel shows the CycleGAN that is trained by real images and simulated annotations with Gaussian blurring. The lower panel shows the CycleGAN that is trained by the same data without Gaussian blurring. The Generator B is used to generate synthetic videos with larger shape variations from circle representations, while the Generator A* generate sharp segmentation for the annotation registrations.}
\label{fig:fig5}
\end{figure}

\begin{figure}
\begin{center}
\includegraphics[width=1\linewidth]{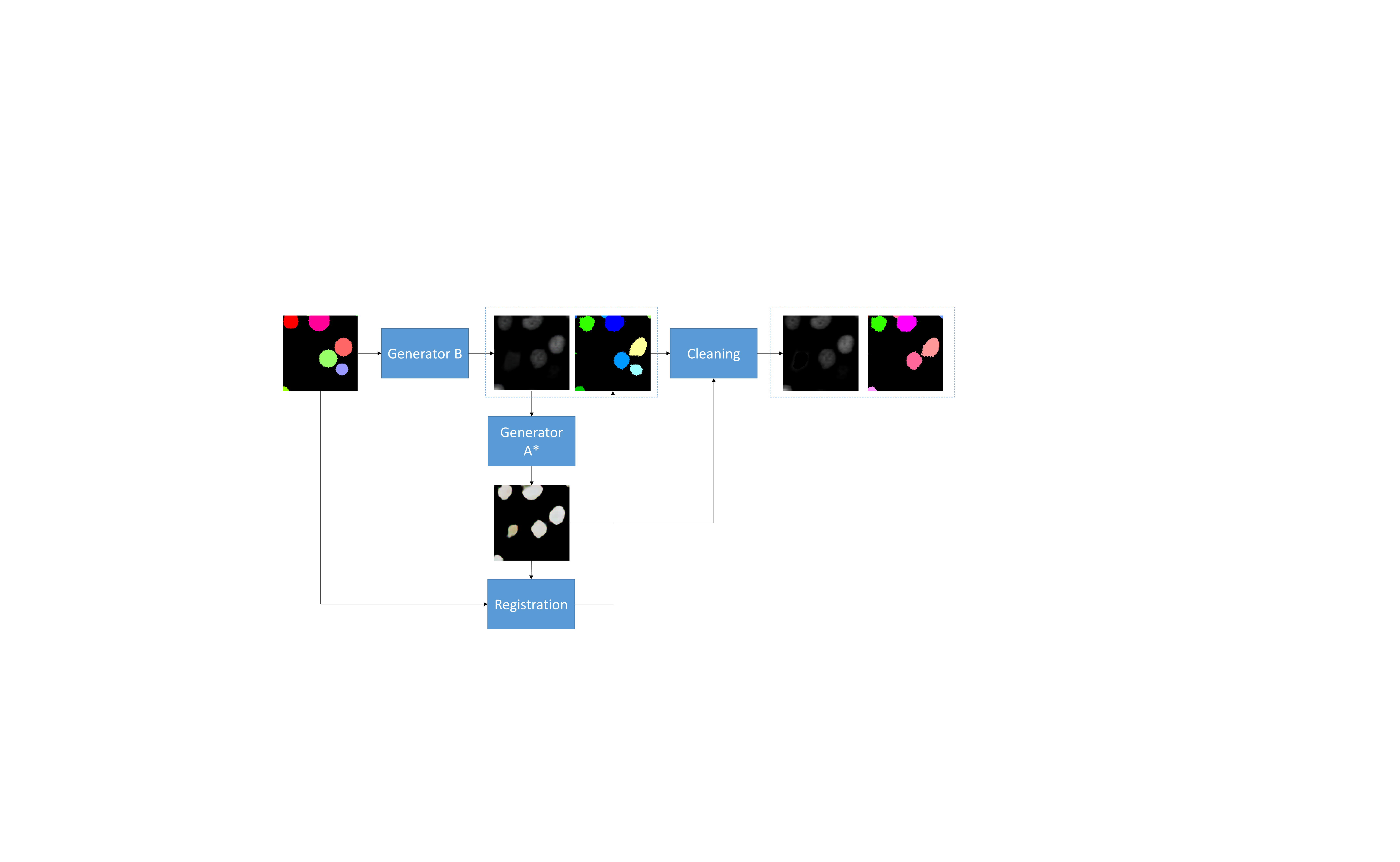}
\end{center}
   \caption{This figures shows the workflow of the annotation refinement approach. The simulated circle annotations are fed into Generator B to synthesize cell images. We used Generator A* in Fig.\ref{fig:fig5} to generate sharp binary masks from synthetic images. Then, we registered simulated circle annotations to binary masks to match the shape of cells in synthetic images. Last, an annotation cleaning step was introduced to delete the inconsistent annotations between deformed instance object masks and binary masks. }
\label{fig:fig6}
\end{figure}

The HeLa cells have higher degrees of freedom in terms of shape variations, compared with microvilli. In this study, we proposed an annotation refinement strategy, to generate shape consistent synthetic HeLa cell videos and annotations, using circles as middle representations (Fig.~\ref{fig:fig5}), without introducing manual annotations. The simulated videos and annotations of HeLa cells are determined by the following operations:

\noindent\textbf{Object number}:The numbers of objects are evaluated when simulating HeLa cell videos. The details are presented in $\mathsection$\textbf{Experimental design}.

\noindent\textbf{Translation}: The instance annotation center can be moved by $N$ pixels. $N$ will be described in $\mathsection$\textbf{Experimental design}.

\noindent\textbf{Radius changing}: Radius of annotations has 10\% probability to get bigger or smaller by 1 pixel.

\noindent\textbf{Disappearing}: Existing instance cells are randomly deleted from certain frames in videos. 

\noindent\textbf{Appearing}: New instance cells shows up from certain frame in videos randomly. New cells will be added to the video from the appearing frame.

\noindent\textbf{Mitosis}: Mitosis is the process of cell replication and splitting. To simulate HeLa cell mitosis, we randomly define "mother cells" at the $n$th frame. At the $n+1$th frame, we delete the "mother cells" and randomly create two new cells nearby. Based on biological knowledge, these two new instances are typically smaller than normal instances, and will grow up bigger and move randomly like other instance annotations. 

\noindent\textbf{Overlapping}: We allow partial overlap between cells. The minimum distance between two cells are set to be 70$\%$ of the total diameter between two cells. 

\noindent\textbf{Size change}: The radius of instance annotation has a 10\% probability to become larger by 1 pixel or become smaller by 1 pixel.

% \begin{figure*}
% \begin{center}
% \includegraphics[width=1\linewidth]{Figures/fig5.pdf}
% \end{center}
%   \caption{This figure shows synthesized mitosis videos and simulated label images. The different panels indicate the different number of simulated objects. Videos in different size are used for different experiment settings.}
% \label{fig:fig5}
% \end{figure*}

% \subsubsection{Video synthesis based on simulated annotations}
% Based on the simulated annotations frames, Generator B trained by CycleGAN will be used to synthesize corresponding microscope videos. 

\subsection{Annotation refinement for HeLa cell video simulation}
After training the initial CycleGAN synthesis, we are able to build simulated videos (with circle representation) as well as their corresponding synthetic microscope videos. However, circles are not the exact shape of annotations for synthetic videos. To further achieve consistent synthetic videos and annotations, we proposed an annotation refinement framework, which has a workflow shown in  Fig.~\ref{fig:fig6}.

\subsubsection{Binary mask generation}

We trained CycleGAN to generate a binary mask of synthetic cell images. Unique from CycleGAN in $\mathsection$\textbf{Unsupervised image-annotation synthesis}, we used training data without applying Gaussian blurring and used the model from an early epoch. From our experiments, we observed that the early epochs of the CycleGAN training focused more on intensity adaptations rather than shape adaptations. The trained Generator A is used to generate sharp binary masks as templates in the following annotation registration step. 

\subsubsection{Annotation deformation (AD)}

To bridge the gap between circle representations and HeLa cell shape annotations, a non-rigid registration approach from ANTs~\cite{avants2011reproducible} is used to deform the circle shapes to the HeLa cell shapes. Briefly, we used generator B to synthesize cell images based on our simulated annotations. In the mask generation, we used generator A* to generate binary masks and registered the circle shape annotations to the binary masks. In that case, we keep the label numbers of circle representations, and deform their shapes to fit the synthetic cells.

\subsubsection{Annotation cleaning (AC)}

When performing image-annotation synthesis using CycleGAN, it is very likely to have a slightly different number of objects between HeLa cell images and annotations without using paired training data. To make the synthetic videos and simulated annotations to have more consistent numbers of objects, we introduce an annotation cleaning step (Fig.~\ref{fig:fig6}). First, we generate binary masks of simulated images using the Generator A*. Second, we clean up the inconsistent objects and annotations by comparing deformed simulated annotations and binary masks. Briefly, pseudovideos are simulated annotations. instance annotations are achieved from binary masks, by treating any connected components as instances. Third, if an instance object in the deformed simulated annotations is not 90\% covered by binary masks, we re-assign the label as background. On the other hand, if a pseudo instance object from the binary masks is not 90\% covered by deformed simulated annotations, we re-assign the corresponding region in the intensity image with the average background intensity values. In sum, the consistent synthetic videos and deformed simulated instance annotations are achieved with annotation cleaning (Fig.~\ref{fig:fig6}).

\subsection{Instance segmentation and tracking}

From the above stages, the synthetic videos and corresponding annotations are achieved frame-by-frame. The next step is to train our instance segmentation and tracking model. We used the recurrent stacked hourglass network (RSHN)~\cite{payer2018instance} as the instance segmentation and tracking backbone to encode the embedding vectors of each pixel. The RSHN is a stacked hourglass network with a convolutional gated recurrent unit to process temporal information. The ideal pixel-embedding has two properties: (1) embedding of pixels belonging to the same objects should be similar across the entire video, and (2) the embedding of pixels belonging to different objects should be different. For a testing video, we employed the Faster Mean-shift algorithm~\cite{zhao2020faster} to cluster pixels to objects as the instance segmentation and tracking results. The embedding-based deep learning methods approach the instance segmentation and tracking as a ”single-stage” approach, which is a simple and generalizable solution across different applications~\cite{payer2018instance,zhao2020faster}.

\section{Experimental Design}
% In this paper, experiments are conducted on two different microscope videos to evaluate the ASIST performance: microvilli videos and HeLa cell videos. The two datasets represent the cellular and sub-cellular microscope videos.

\subsection{Instance segmentation and tracking on microvilli video}

\subsubsection{Data}
Two microvilli videos captured by fluorescence microscopy are in 1.1$\mu$m pixel resolution. Training data is one microvilli video in 512$\times$512 in pixel resolution. Testing data is another microvilli video in the size of 328$\times$238 pixels. Due to the heavy load of manual annotations on video frames, we only annotated the first ten frames of both videos as the golden standard. The annotation work includes two parts: 1) first we annotated each microvilli structure including overlapping or densely distributed areas; 2) secondly, each instance has been assigned consistent labels across all frames in same video. The manual annotation labor on both training and testing data takes roughly a week of work from a graduate student. This long manual annotation process shows the value of annotation-free solutions in quantifying cellular and subcellular dynamics.

\subsubsection{Experimental design}
In order to assess the performance of our annotation-free instance segmentation and tracking model, the proposed method is compared with the model trained with manual annotations on the same testing microvilli video. The different experimental settings are shown as the following: 

\noindent\textbf{Self}: The testing video with manual annotations was used as both training and testing data. 

\noindent\textbf{Real}: Another real microvilli video with manual annotations were used as training data.

% \noindent\textbf{Microvilli-1 10 frames}: training data used 1 simulated video contains 100 instances in size of 512$\times$512. Video is simulated as described in $\mathsection$\textbf{Method}. Only first 10 frames are used in model training.

\noindent\textbf{Microvilli-1}: One simulated video which consisted of 100 instances in size of 512$\times$512 pixels was used as training data. The "Microvilli-1 10 frames" indicated only 10 frames were used, while other simulated data used 50 frames.

\noindent\textbf{Microvilli-5}: Five simulated videos with 512$\times$512 pixel resolutions were used as training data. The number of objects were empirically chosen to be between 80 to 220. 

\noindent\textbf{Microvilli-20}: We further spatially split each 512$\times$512 video in Microvilli-5 to four 256$\times$256 videos to form a total of 20 simulated videos with half resolution.

\subsection{Instance segmentation and tracking on HeLa cell video}

\subsubsection{Data}
HeLa cell videos (N2DL-HeLa) were obtained from the ISBI Cell Tracking Challenge~\cite{mavska2014benchmark,ulman2017objective}. The cohort has two 92-frame HeLa cell videos in size of 1100$\times$700 pixels with annotations. The second video with complete manual annotations is used as the testing data for all experiments.

\subsubsection{Experimental design}
For experiments using an annotation-free framework, synthetic videos and simulated annotations are used for training. As a comparison experiment, experiments trained with annotated data used two N2DL-HeLa videos with annotations as training data. Our experiment settings are described as follows:

\noindent\textbf{Self}: The testing video with manual annotations was used as both training and testing data. The patch size of 256$\times$256 was used, following~\cite{payer2018instance,zhao2020faster}. 

\noindent\textbf{Self-HW}: The testing video with manual annotations was used as both training and testing data. The patch size of 128$\times$128 was used, as a half window (HW) size. 

\noindent\textbf{HeLa}: Our training data was 10 simulated videos with 512$\times$512 resolution containing approximately 150 objects, including 20 cell appearing events, 20 cell disappearing events, and 5 or 10 mitosis events. The numbers were empirically chosen. This experiment employed the circle annotations directly as the baseline performance. The patch size of 256$\times$256 was used.

\noindent\textbf{HeLa-AD}: The above simulated data was used for training, with an extra annotation deformation (AD) step. 

\noindent\textbf{HeLa-AD+AC}: The above simulated data was used for training, with extra AD and annotation cleaning (AC) steps. 

\noindent\textbf{HeLa-AD+AC+HW}: The above simulated data was used for training, with extra AD and AC steps. The patch size of 128$\times$128 was used, as a half window (HW) size. 

\subsection{Evaluation matrix}

TRA, DET, and SEG are the standard metrics in the ISBI cell tracking challenge~\cite{matula2015cell}, evaluating the performance of tracking, detection and segmentation, respectively. The ISBI Cell Tracking Challenge used these three metrics as \textit{de facto} standard measurements based on Acyclic Oriented Graph Matching (AOGM) algorithms. The instance objects are presented as the nodes of the acyclic oriented graphs, while the tracking results are modeled as the vertices of the graphs. Then, graphs are obtained from both ground truth annotations and the predicted results to evaluate the accuracy of detection (DET) and tracking (TRA). SEG evaluates the overlap of predicted objects with true objects. The TRA, DET and SEG range from 0 to 1, where 0 and 1 indicate the worst and best performance, respectively. The details of such metrics can be found in~\cite{matula2015cell}.

\section{Results}

\subsection{Instance segmentation and tracking on microvilli videos}
\begin{figure*}
\begin{center}
\includegraphics[width=0.88\linewidth]{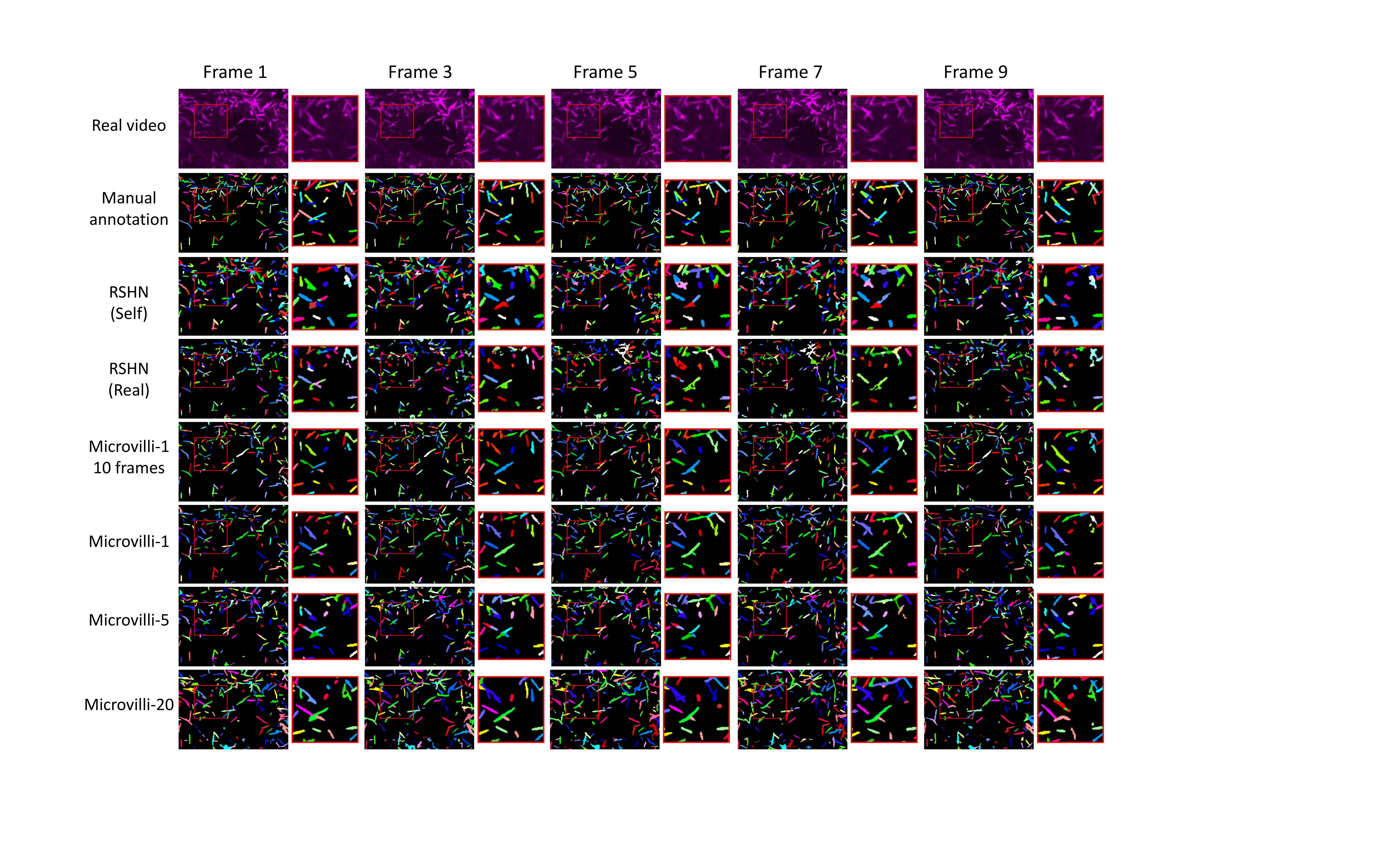}
\end{center}
   \caption{This figure shows the instance segmentation and tracking results of the real testing microvilli video.}
\label{fig:fig7}
 \end{figure*}

The qualitative and quantitative results are presented in Fig.~\ref{fig:fig7} and Table.~\ref{tab:TRA1}. From the quantitative results shown in Table.~\ref{tab:TRA1}, the best performance according to the evaluation metric scores was achieved by Microvilli-20 without using manual annotations. By contrast,it took one week of manual annotation labor from a graduate student to annotate only 10 frames of RSHN (Self) and RSHN (Real). One salient feature of achieving better performance of the proposed framework is the larger number of total simulated training video.

\begin{table}[]
\caption{DET, SET and TRA values of different experiments on microvilli video.}
\centering
\begin{tabular}{c|c|c|ccc}
\hline
Exp. & T.V. & T.F. & DET & SEG & TRA \\
\hline
RSHN (Self)~\cite{payer2018instance}& 1 & 10 & 0.662 & 0.298 & 0.629 \\
\hline
RSHN (Real)~\cite{payer2018instance}& 1 & 10 & 0.357 & 0.169 & 0.334 \\
ASIST (Microvilli-1) & 1 & 10 & 0.580 & 0.306 & 0.551 \\
\hline
ASIST (Microvilli-1) & 1 & 50 & 0.586 & 0.311 & 0.556 \\
ASIST (Microvilli-5) & 5 & 50 & 0.660 & \textbf{0.338} & 0.627 \\
ASIST (Microvilli-20) & 20 & 50 & \textbf{0.715} & 0.332 &  \textbf{0.674} \\
\hline
\end{tabular}
\noindent 
T.V. is the number of training videos. T.F. is the number of training frames of each video. RSHN (Self) uses testing video for training. RSHN (Real) is the standard testing accuracy of using another independent video as training data.
\label{tab:TRA1}
\end{table}

\subsection{Instance segmentation and tracking on HeLa cell videos}

\begin{figure*}
\begin{center}
\includegraphics[width=0.88\linewidth]{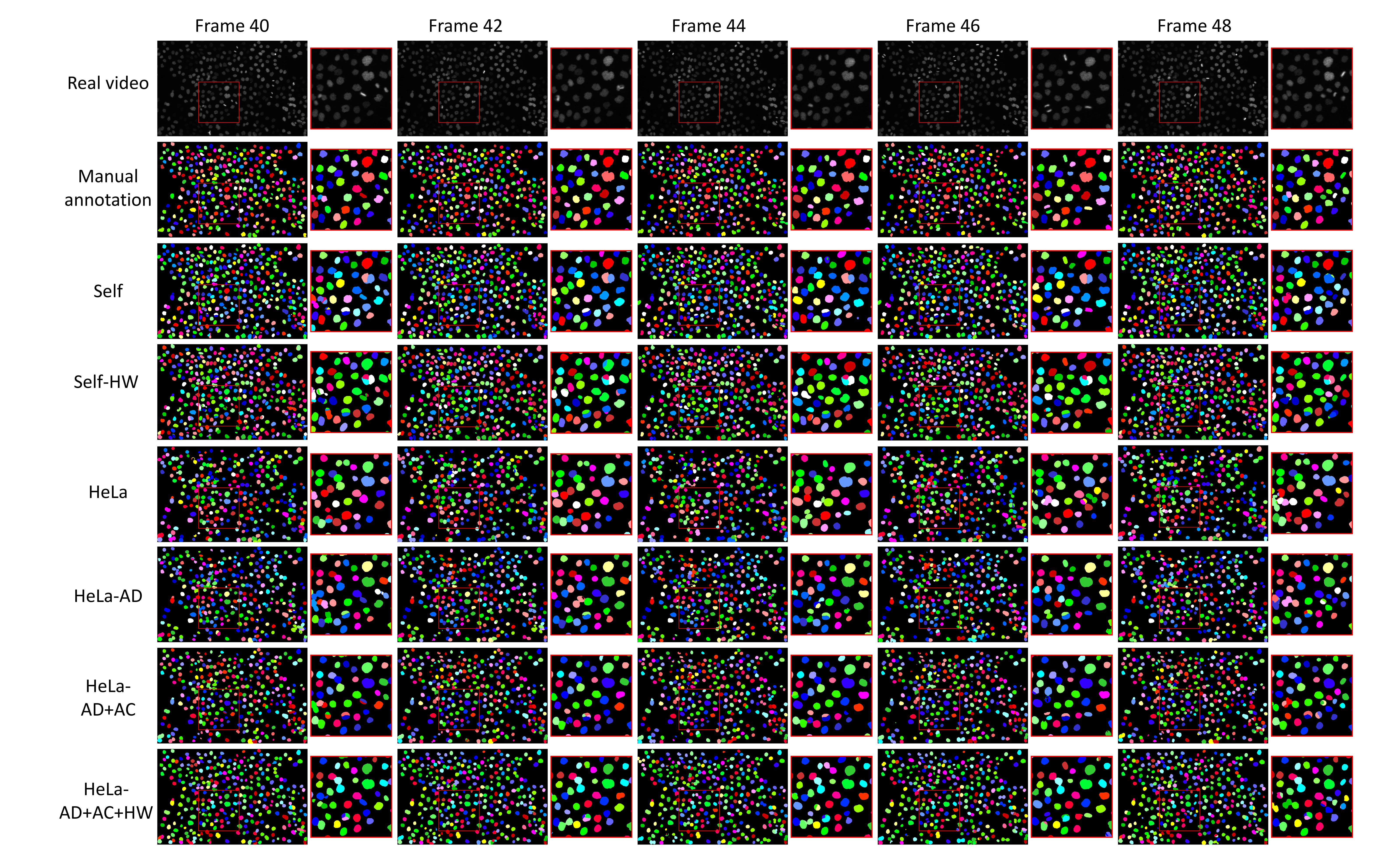}
    \caption{This figure shows the instance segmentation and tracking results on the real HeLa cell testing video.}
\end{center}
\label{fig:fig8}
\end{figure*}

Instance segmentation and tracking results of HeLa cell videos were presented in Fig.~\ref{fig:fig8}. Based on the performance in Table.~\ref{tab:TRA2}. HeLa-AD+AC+HW achieved superior performance than other settings using the ASIST method. The best performance of our annotation-free ASIST method is 5\% to 9\% lower than the manual annotation baseline. The most salient feature of improving the performance is to introduce the annotation cleaning (AC) step.

\begin{table}[]
\caption{DET, SET and TRA values of different experiments on HeLa cell video.}
\centering
\begin{tabular}{c|c|c|ccc}
\hline
Exp. & T.V. & T.F. & DET & SEG & TRA \\
\hline
RSHN (Self)~\cite{payer2018instance} & 2 & 92  & \textbf{0.979} & \textbf{0.884} & \textbf{0.975} \\
RSHN (Self-HW) & 2 & 92  & 0.956 & 0.809 & 0.951 \\
\hline
ASIST (HeLa) & 10 & 50 & 0.858 & 0.656 & 0.849 \\
ASIST (HeLa-AD) & 10 & 50 & 0.853 & 0.718 & 0.844 \\
ASIST (HeLa-AD+AC) & 10 & 50 & 0.919 & 0.755 & 0.911 \\
ASIST (HeLa-AD+AC+HW) & 10 & 50 & 0.939 & 0.796 & 0.928 \\
\hline
\end{tabular}
\noindent 
T.V. is the number of training videos. T.F. is the number of training frames per video. RSHN (Self) is the upper bound of RSHN using testing video for training.
\label{tab:TRA2}
\end{table}

% \subsection{Instance segmentation and tracking on HeLa cell videos}
% 1. In the experiment on microvilli videos, our ASIST method, ASIST (Microvilli-20), performs best.

% 2. In the experiment on Hela cell videos, the best performance of the ASIST method is 5\% to 9\% lower than manual annotation baseline.

\section{Discussion}
In this paper, we assess the feasibility of performing pixel-embedding based instance object segmentation and tracking in an annotation-free manner, with adversarial simulations. Compared with conventional segmentation and tracking methods on microscope videos, our experiment used a pixel-embedding strategy instead of the “segmentation and association” two-step method. Our method also used synthetic training data instead of manual annotation. According to our experimental results, our annotation-free instance segmentation and tracking model achieved superior performance on the microvilli dataset as well as comparable results on the HeLa dataset. Such encouraging results elucidated a promising new path to leverage the currently unsalable human annotation based pixel-embedding deep learning approach in an annotation free manner. In terms of robustness, the proposed pixel-embedding based method does not require heavy parameter tuning, which is typically inevitable in traditional model based methods. As a learning based method, the robustness of the proposed method can be further improved with more heterogeneous training images. 

\textbf{Strengths. } The strength of our proposed ASIST method is three-fold: I. the proposed method is annotation-free to alleviate the extensive manual efforts of preparing large-scale manual annotations for training deep learning approaches; II. The proposed method does not require heavy parameter tuning; III. The proposed ASIST method combines the strength of both adversarial learning and pixel embedding based cell instance segmentation and tracking.

\textbf{Limitations. } One major limitation of our ASIST method is that both microvilli and HeLa cells have relatively homogeneous shape and appearance variations. In the future, it will be valuable to explore more complicated cell lines and more heterogeneous microscope videos. Meanwhile, the registration based method is introduced to capture the shape variations for ball-shaped HeLa cells. For more complicated cellular and subcellular objects, deep learning based solutions might be needed, such as the shape auto-encoder.

Following the proposed ASIST framework, our long term goal is to propose more general and comprehensive algorithms that can be applied to a variety of microscope videos with pixel-level instance segmentation and tracking. This would provide new analytical tools for domain experts to characterize high spatio-temporal dynamics of cells and subcellular structures.

\section{Conclusion}
In this paper, we propose the ASIST method – an annotation-free instance segmentation and tracking solution to characterize cellular and subcellular dynamics in microscope videos. Our method consists of unsupervised image-annotation synthesis, video synthesis, and instance segmentation and tracking. According to the experiments on subcellular (microvilli) videos and cellular (HeLa cell) videos, ASIST achieved comparable performance to manual annotation-based strategies. The proposed approach is a novel step towards annotation-free quantification of cellular and subcellular dynamics for microscope biology.

% \section*{Acknowledgment}

\ifCLASSOPTIONcaptionsoff
  \newpage
\fi



\bibliographystyle{IEEEtran}
\bibliography{main}


% \end{thebibliography}


%\vfill


\end{document}